%% file: main.tex
\documentclass[sigconf]{acmart}

\usepackage{flushend}
\usepackage{balance}

\def\BibTeX{{\rm B\kern-.05em{\sc i\kern-.025em b}\kern-.08emT\kern-.1667em\lower.7ex\hbox{E}\kern-.125emX}}

\acmSubmissionID{XXXXXX}

\usepackage{multirow}
\usepackage{booktabs}
\usepackage{graphicx}
\usepackage{subfigure} 
\usepackage{caption}
\usepackage{url}       
\usepackage{stfloats}  
\usepackage{array}
\usepackage{breqn}
\usepackage{amsfonts}
\usepackage{courier}

\usepackage{amssymb}
\usepackage{balance}
\usepackage{multirow,tabularx}
\usepackage{longtable}
\usepackage{booktabs,longtable}
\usepackage{hhline}
\usepackage{dsfont}
\usepackage{fancyhdr}
\usepackage{epstopdf}
\usepackage{algorithm}
\usepackage{enumitem}
\usepackage{algorithmic}
\usepackage[bottom]{footmisc}
\usepackage{verbatim}

\newcommand{\etal}{et al.}

\newcolumntype{L}[1]{>{\raggedright\let\newline\\\arraybackslash\hspace{0pt}}m{#1}}
\newcolumntype{C}[1]{>{\centering\let\newline\\\arraybackslash\hspace{0pt}}m{#1}}
\newcolumntype{R}[1]{>{\raggedleft\let\newline\\\arraybackslash\hspace{0pt}}m{#1}}

\newfont{\mycrnotice}{ptmr8t at 7pt}
\newfont{\myconfname}{ptmri8t at 7pt}

\copyrightyear{2023}
\acmYear{2023}
\setcopyright{acmlicensed}\acmConference[SIGIR '23]{Proceedings of the 46th International ACM SIGIR Conference on Research and Development in Information Retrieval}{July 23--27, 2023}{Taipei, Taiwan}
\acmBooktitle{Proceedings of the 46th International ACM SIGIR Conference on Research and Development in Information Retrieval (SIGIR '23), July 23--27, 2023, Taipei, Taiwan}
\acmPrice{15.00}
\acmDOI{10.1145/3539618.3591864}
\acmISBN{978-1-4503-9408-6/23/07}

\begin{document}
\fancyhead{}
\title{OFAR: A Multimodal Evidence Retrieval Framework for Illegal Live-streaming Identification}

\author{Dengtian Lin}
\affiliation{%
	\institution{Shandong University}
}\email{lindengtian@mail.sdu.edu.cn}

\author{Yang Ma}
\affiliation{%
	\institution{Alibaba Group}
}\email{mayang.ma@alibaba-inc.com}

\author{Yuhong Li}
\affiliation{%
	\institution{Alibaba Group}
}\email{daniel.lyh@alibaba-inc.com}

\author{Xuemeng Song}
\affiliation{%
	\institution{Shandong University}
}\email{sxmustc@gmail.com}

\author{Jianlong Wu}
\affiliation{%
	\institution{Shandong University}
}\email{jlwu1992@pku.edu.cn}

\author{Liqiang Nie}
\affiliation{%
	\institution{Shandong University}
}\email{nieliqiang@gmail.com}
\renewcommand{\shortauthors}{Dengtian Lin et al.}


\begin{abstract}
Illegal live-streaming identification, which aims to help live-streaming platforms immediately recognize the illegal behaviors in the live-streaming, such as selling precious and endangered animals, plays a crucial role in purifying the network environment. %
Traditionally, the live-streaming platform needs to employ some professionals to manually identify the potential illegal live-streaming. Specifically, the professional needs to search for related evidence from a large-scale knowledge database for evaluating whether a given live-streaming clip contains illegal behavior, which is time-consuming and laborious. 
To address this issue, in this work, we propose a multimodal evidence retrieval system, named OFAR, to facilitate the illegal live-streaming identification. OFAR consists of three modules: \textit{Query Encoder}, \textit{Document Encoder}, and \textit{MaxSim-based Contrastive Late Intersection}. 
Both query encoder and document encoder are implemented with the advanced \mbox{OFA} encoder, which is pretrained on a large-scale multimodal dataset. 
In the last module, we introduce contrastive learning on the basis of the MaxiSim-based late intersection, to enhance the model's ability of query-document matching. 
The proposed framework achieves significant improvement on our industrial dataset TaoLive, demonstrating the advances of our scheme. 

\end{abstract}

\fancyhf{}



\maketitle
\input{sec-intro}
\input{sec-rel}
\input{sec-model}
\input{sec_exp}

\input{sec-con}

\begin{acks}
This work is supported by the National Natural Science Foundation of China Grant, No.: U1936203, No.: 62006140 and No.: 62236003; 
the Shandong Provincial Natural Science Foundation, No.: ZR2022YQ59; 
the Alibaba Group through Alibaba Innovative Research Program; 
and the Shenzhen College Stability Support Plan Grant, No.: GXWD20220817144428005.
\end{acks}

\clearpage
\section*{Presenter and Company Biography}
Dengtian Lin is an Algorithm Intern at Alibaba Group, where he is involved mainly in information retrieval systems. Dengtian Lin is also a graduate student in the School of Computer Science and Technology, Shandong University.

Alibaba Group Holding Ltd (Alibaba Group) is a provider of e-commerce and technology infrastructure services. The company provides fundamental technology infrastructure services to merchants, brands, retailers, and businesses to market, sell and operate using the Internet. Its businesses comprise core commerce, digital media and entertainment, cloud computing, and other innovation initiatives. Its mission is to make it easy to do business anywhere.

\bibliographystyle{ACM-Reference-Format}  
\balance
\bibliography{sigproc_abbre}

\end{document}

%% file: sec-intro.tex
\section{Introduction}
\label{sec:intro}
In recent years, with the advances in multimedia technology, live e-commerce has become a major sales channel for many brands and companies. The TaoLive platform is one of the largest live-streaming platforms for e-commerce in China, where there are more than $500,000$ live streams in 2022\footnote{\url{https://static.alibabagroup.com/reports/fy2022/ar/ebook/en/26/index.html}}. 
As the live-streaming allows sellers to freely broadcast their products, there may contain certain illegal behaviors, like selling precious and endangered animals, or misrepresentation of product effectiveness. 
Therefore, live e-commerce platforms usually need to employ professionals to identify and publish potential illegal live-streaming. 
Typically, if the professionals consider that the current live-streaming clip probably contains illegal behavior, they would manually search for related evidence from the database to identify the illegal behavior. 
Obviously, this is time-consuming and laborious for browsing all the live-streams. Therefore, it is highly desirable to develop an automatic evidence retrieval method for illegal live-streaming identification. 

\begin{figure}
     \centering
     \includegraphics[width=0.45\textwidth]{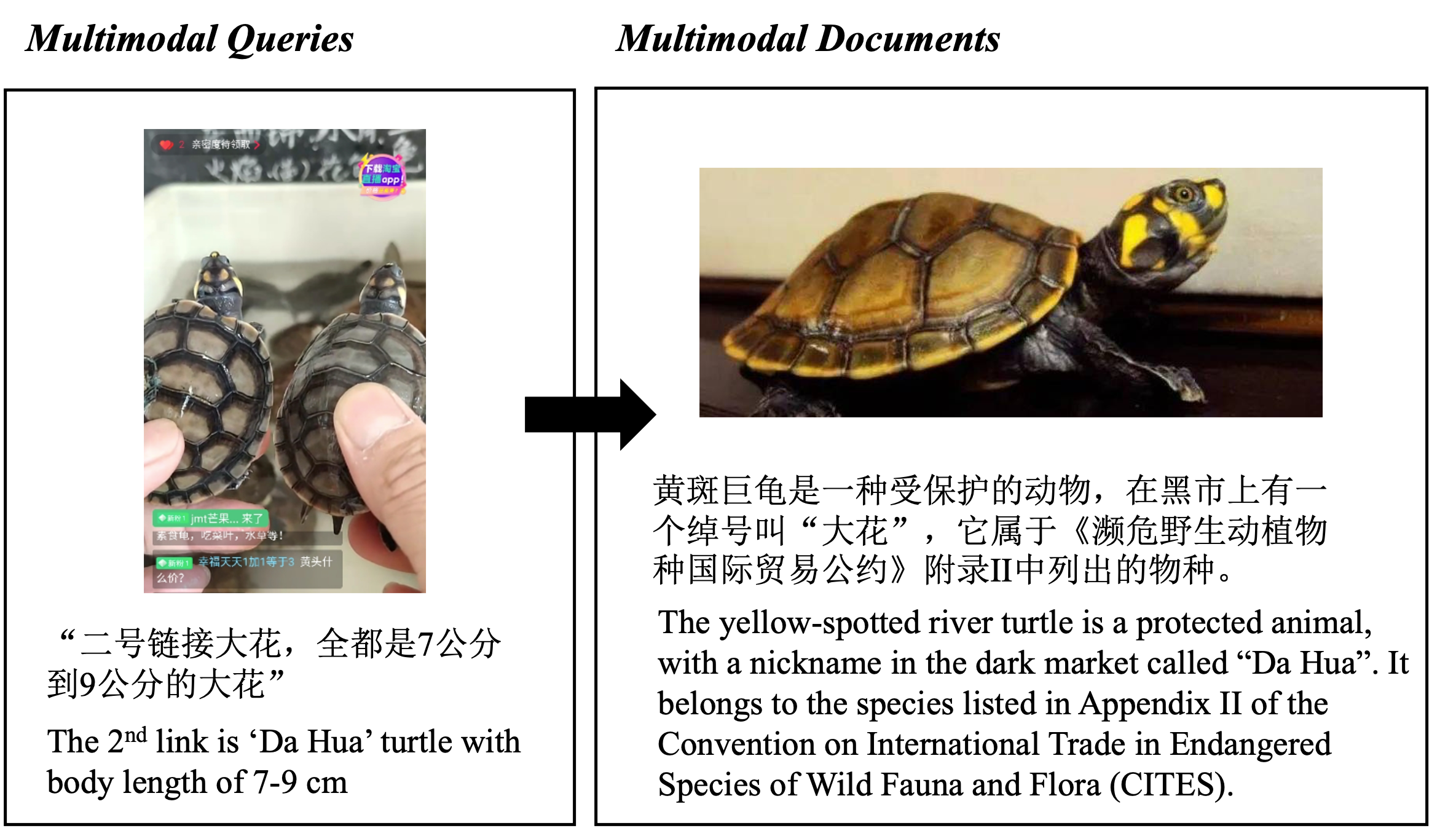}
     \caption{A multimodal evidence retrieval case for our TaoLive platform. The context is that a host is illegally selling endangered turtles through live-streaming, which should be quickly identified and punished by the platform. Then retrieving multimodal evidence documents to support the illegal live-streaming identification.}
     \label{fig:intro_cases}
\end{figure}

Figure~\ref{fig:intro_cases} illustrates an application case, where the seller is advertising an endangered species, whose nickname is ``Da Hua''. The automatic evidence retrieval method is expected to retrieve the related evidence (documents) to facilitate the professional to judge whether the given live-streaming screenshot and the corresponding transcripts derived by conducting Automatic Speech Recognition~(ASR) over the live speech contain illegal  behavior. 
Essentially, this task can be regarded as the information retrieval~(IR) task, which aims to retrieve relevant documents for a given query.  

In fact, with the prevalence of neural networks, many neural IR models have emerged. Early neural IR models~\cite{DBLP:journals/access/ChenZZY20, DBLP:conf/sigir/MacAvaneyYCG19, DBLP:conf/naacl/YangXLLTXLL19} mainly first concatenate the query and document, and then adopt a single encoder with a cross-attention mechanism to capture their interactions. 
Despite achieving remarkable performance, these methods come with great computation costs, and thus cannot be applied in large-scale retrieval applications. Therefore, certain researchers have proposed the dual-attention based models, which employ two encoders to separately encode queries and documents and utilize a dual-attention mechanism to compute their relevance score, to speed up the retrieval process. 
For example, 
RocketQA~\cite{qu2021rocketqa} employs two separate encoders to encode the query and document, respectively, both of which are initialized with the same pretrained language model~(\textit{i.e.}, ERNIE~\cite{DBLP:journals/corr/abs-1904-09223}), and using dot-production based attention mechanism on evaluating the relevance score between the query and document.
In addition, ColBERT~\cite{khattab2020colbert} utilizes two BERT-based~\cite{devlin2019bert} encoders to obtain token-level representations of the query and document. And then ColBERT adopts a maximum similarity~(MaxSim) based late intersection mechanism to evaluate the relevance score between queries, which is a token-level attention mechanism to evaluate their relevance score between queries and documents using the inner product  

Inspired by these IR models, for our context, we design a multimodal evidence retrieval system to facilitate the illegal live-streaming identification. 
As illustrated in Figure~\ref{fig:model_figure}, our proposed system, named OFAR, consists of three modules: \textit{Query Encoder}, \textit{Document Encoder}, and \textit{MaxSim-based Contrastive Late Intersection}. 
Notably, both query encoder and document encoder are implemented with the advanced \mbox{OFA}~\cite{wang2022ofa} encoder,  which can better handle multimodal data with a unified encoder. 
In the last module, we introduce an innovative MaxSim-based contrastive late intersection, to complete the multimodal similarity computation efficiently. 
Different from ColBERT which only involves one negative document with the triple loss for query-document matching learning, we adopt contrastive learning on the basis of the MaxiSim-based late intersection, to incorporate more negative documents, and hence improve our model's ability of query-document matching.  Ultimately, to promote the retrieval performance of our model, we first pretrain our model on two open-source IR datasets, namely WebQA~\cite{chang2022webqa} and \mbox{MS MARCO}~\cite{DBLP:conf/nips/NguyenRSGTMD16}, and then finetune it on our internal training dataset TaoLive. 

Our main contributions can be summarized as follows:
\begin{itemize}
\item We propose a multimodal evidence retrieval system OFAR for facilitating the illegal live-streaming identification. In particular, we propose an innovative MaxSim-based contrastive late intersection mechanism, which can realize effective and efficient evidence retrieval.  
\item Our experiments show that OFAR achieves significant improvement over OFAR without MaxSim on our internal dataset. The proposed method also improves our consensus passages retrieval task by 59.8\% at most.  
\end{itemize}

%% file: sec-rel.tex
\section{Related Work}
\label{sec:rel}
In this section, we give the related work on neural information retrieval models, which is the core technology for solving our illegal live-streaming identification task. 

Neural information retrieval models aim to retrieve the relevant context from a large corpus given a query with advanced neural networks.
Existing models can be mainly classified into two groups: cross-attention based models~\cite{DBLP:journals/access/ChenZZY20, DBLP:conf/sigir/MacAvaneyYCG19, DBLP:conf/naacl/YangXLLTXLL19}, and dual-attention based models~\cite{DBLP:conf/emnlp/KarpukhinOMLWEC20, DBLP:conf/iclr/ChangYCYK20, qu2021rocketqa, DBLP:conf/emnlp/RenQLZSWWW21, khattab2020colbert, DBLP:conf/naacl/SanthanamKSPZ22}. 
The cross-attention based models first concatenate the query and document, and then utilize a single encoder equipped with the cross-attention mechanism to learn the latent interaction between the query and document. Typically, the cross-attention based models can achieve remarkable retrieval performance but suffer from heavy computation cost and hence is not suitable for retrieval in the large-scale corpus.
On the contrary, the dual-attention based models first employ separate encoders to encode the queries and documents, respectively, and then utilize dual-attention mechanism to fulfill the query-document matching. 
For example, 
\mbox{Humeau~\etal~\cite{DBLP:conf/iclr/HumeauSLW20}} proposed a new Poly-encoder based on the Transformer architecture for encoding the queries and documents, respectively, and uses the dual-attention mechanism for learn the query-document relevance. 
Later, Khattab and Zaharia~\cite{khattab2020colbert} adopted the dual encoder architecture and introduced a new dual-attention mechanism, named MaxSim-based late interaction, for efficiently conducting the query-document matching. 
One advantage of the dual-attention based models is that they support the offline pre-computation of document representations and leverage certain advanced efficient search mechanisms, such as MIPS~\cite{DBLP:conf/nips/Shrivastava014} and faiss~\cite{johnson2019billion}, during the inference phase.

Although these dual-attention based models have achieved great success in speeding up the retrieval process, they generally use a triple loss to optimize query-document matching learning. In fact, the triple loss only involves one negative document, causing sub-optimal retrieval performance. To address this issue, we introduce the contrastive learning to improve our model's ability of query-document matching. 

%% file: sec-model.tex
\section{Proposed Framework}
\label{sec:meth}

\begin{figure*}[!t]
    \centering
    \includegraphics[scale=0.52]{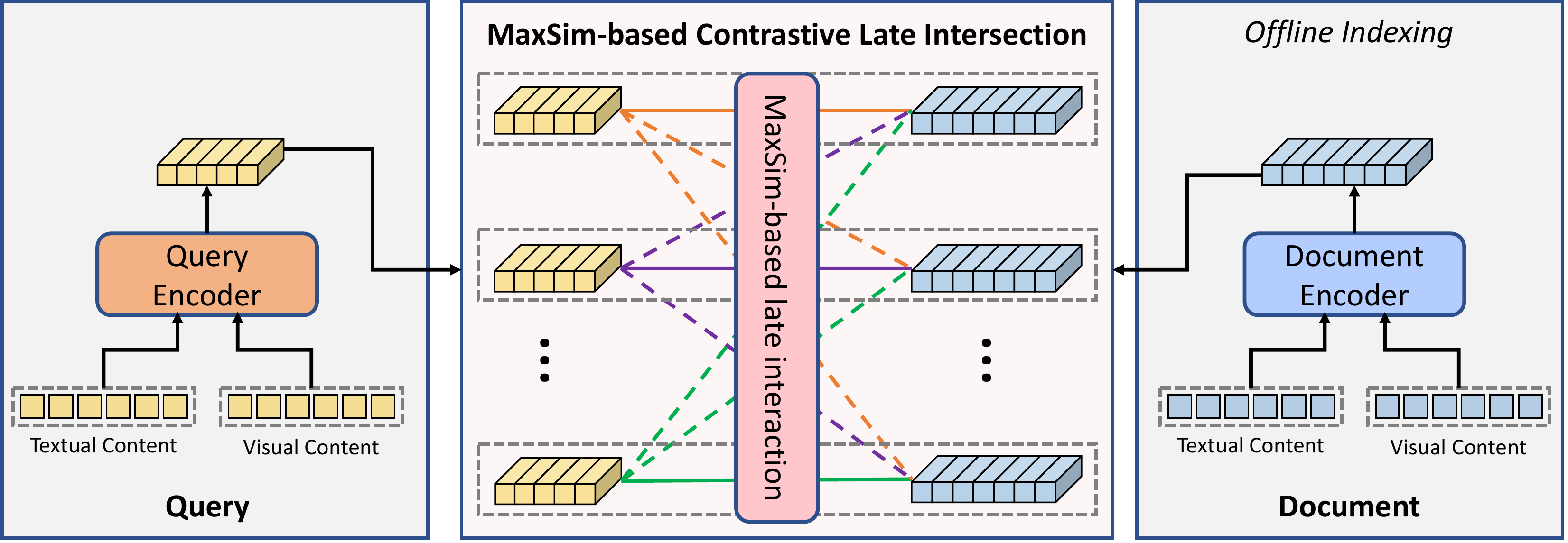}
    \caption{The proposed model OFAR, which consists of three vital modules: \textit{Query Encoder}, \textit{Document Encoder}, and \textit{MaxSim-based Contrastive Late Intersection}.}
    \label{fig:model_figure}
\end{figure*}



The proposed model OFAR consists of three vital modules: \textit{Query Encoder}, \textit{Document Encoder}, and \textit{MaxSim-based Contrastive Late Intersection}. 

\subsection{Query/Document Encoder}
\label{subsec:QD_encoders}
In this work, both query encoder and document encoder are implemented with the encoder of the multimodal pretrained model OFA, which has shown compelling success in many multimodal tasks~\cite{wang2022ofa,tang2022you}.

\textbf{Query Encoder.}  
According to the OFA encoder, the textual content of the query will first be tokenized by the byte-pair encoding method of BERT~\cite{devlin2019bert}. 
The visual content of the query will be first encoded into patch features by a ResNet module and then projected into high-level visual tokens by the image quantization technology~\cite{DBLP:conf/cvpr/EsserRO21} based on the extracted patch features. 
Thereafter, the textual tokens and visual tokens will be concatenated as a unified sequence and processed through several Transformer blocks and a linear layer with L2 normalization. Notably, in practice, the unified sequence is padded to a predefined length. 
Formally, we have 
\begin{equation}
    \mathbf{E}_q = \mathcal{F}_Q(q),
\end{equation}
where {$\mathbf{E}_q=[\mathbf{e}^1_q;\mathbf{e}^2_q;\cdots;\mathbf{e}^{M_q}_q]\in \mathbb{R}^{M_q\times D_E}$} is the representation of the query $q$, and  $\mathcal{F}_Q$ is the OFA-based query encoder. $M_q$ refers to the length of the query $q$. $D_E$ represents the dimension of the representation. 

\textbf{Document Encoder.}  
Similarly, for encoding each document, we adopt a separate OFA encoder, denoted as $\mathcal{F}_D$.
Mathematically, we have 
\begin{equation}
    \mathbf{E}_d = \mathcal{F}_D(d),
\end{equation}
where $\mathbf{E}_d=[\mathbf{e}^1_d;\mathbf{e}^2_d;\cdots;\mathbf{e}^{M_d}_d] \in \mathbb{R}^{M_d\times D_E}$ is the representation of document $d$. $M_d$ refers to the length of the document $d$. 


\subsection{MaxSim-based Contrastive Late Intersection}
\label{subsec:CLI}

Motivated by the powerful capability of the MaxSim-based late intersection mechanism introduced by ColBERT~\cite{khattab2020colbert} in promoting the retrieval performance and speeding up the retrieval process,  we resort to the MaxSim-based late intersection as the basis for the query-document matching. 
In particular, we introduce contrastive learning~\cite{V2P, CL_work1, CL_work2, CL_work3}, which aims to make similar (positive) samples closer and separate dissimilar (negative) samples far from each other,  to further promote the retrieval performance. 

Specifically, according to the MaxSim-based late intersection, the relevance score between the query $q$ and document $d$ can be measured as follows,
\begin{equation}
\mathbf{S}_{q,d}=\sum_{i}^{M_q}\max_{j}^{M_d}\mathbf{E}_{q_i} \mathbf{E}^T_{d_j},
\label{equ:Sqd}
\end{equation}
where $\mathbf{S}_{q,d}$ denotes the relevance score of the document $d$ for the query $q$. Intuitively, MaxSim-based late intersection decomposes the relevance score of the document for the query into multiple token-wise relevance scores of the document for the query. To be specific, for each token in the query, we calculate its inner product with each token in  the document and then adopt the maximum value as the overall relevance score of the document towards the given token. Ultimately, by summing all the token-wise relevance scores of the document for the query, we can derive the final relevance score of the document for the given query.


Different from ColBERT, which uses the triple loss that only involves a single negative document, we adopt the contrastive learning loss to incorporate more negative documents, to enhance the model's ability on distinguishing positive and negative documents. 
In particular, we randomly sample a batch of $B$ training query-document pairs $\{(q^b, d^b)\}_{b=1}^B$. We regard each pair $(q^b, d^b)$ as a positive pair, while all the pairs $(q^u, d^b)$, where $b\neq u$, are the corresponding negative pairs. 
Finally, we deploy the following contrastive loss for model optimization,
\begin{equation}
    \mathcal{L}_{CL}=-\log \sum_{u=1}^{B}{\frac{exp({\mathbf{S}_{q^u,d^u}/\tau)}}{\sum_{b=1}^{B}{\mathbf{1}_{[b\neq u]}exp(\mathbf{S}_{q^u,d^b}/\tau)}}},
    \label{equ:contrastive}
\end{equation}
where $\mathbf{1}_{[b\neq u]} \in \{0,1\}$ is an element-wise indicator function, which equals to $1$ if $b\neq u$, otherwise $0$. $\tau$ represents the positive temperature hyperparameter. 


%% file: sec_exp.tex
\section{Experiment}
\label{sec:exp}
In this section, we present the experimental setting as well as the performance of our model.
\subsection{Experimental Setting}
\subsubsection{Dataset.}
To evaluate our proposed model, we used an internal dataset of our TaoLive platform, which contains more than $55K$ <query, evidence document> pairs. Each pair corresponds to an illegal case that has been identified by the professional.
In each pair, the query is composed of a screenshot of a live-streaming clip that has been identified as an illegal case by the TaoLive platform, and its corresponding transcript extracted by ASR. The corresponding evidence document is selected by the professional from an internal large-scale multimodal knowledge database, which is daily updated by collecting the related regulations\footnote{\url{http://www.forestry.gov.cn/main/5461/20210205/122418860831352.html}} from several government official websites, and related news published by official media\footnote{\url{https://weibo.com/3277148477/LqmgSkjtx?refer_flag=1001030103_}}. Our TaoLive dataset is split into a training set with $50,504$ samples and a test set with $5,688$ samples. On average, the length of the query and the evidence document is $248$ and $260$, respectively. 


Considering the scale of our dataset, we 
introduced two public information retrieval datasets: a textual dataset \mbox{MS MARCO}~\cite{DBLP:conf/nips/NguyenRSGTMD16} and a multimodal dataset WebQA~\cite{chang2022webqa}, to pretrain our model and enhance its query-document matching capability.
\mbox{MS MARCO} contains more than $8,800K$ passages from Web pages as source documents, where queries are associated with these documents.
The WebQA dataset has over $46K$ question-answer pairs, where each question is associated with not only an answer but also a relevant document to reason the answer.
Finally, for pretraining, we merge these two datasets together. Considering their heterogeneity, we set the visual content with a blank image if it is missing in the query or document. If the textual content is missing, we directly feed the visual content into the query/document encoder.  

\subsubsection{Implementation Details.} 
We used the OFA\footnote{\url{https://github.com/OFA-Sys/OFA}} encoder as the backbone of our Query/Document encoder. 
For both pretraining on WebQA and \mbox{MS MARCO}, and finetuning on the TaoLive dataset, the learning rate for Adam optimization was set to $1e$-$4$. The batch size $B$ was set to $32$. The dimension of embedding representation $D_E$ was set to $768$. The temperature hyperparameter $\tau$ is set to $1.0$. 
Similar to previous works~\cite{khattab2020colbert, DBLP:conf/sigir/WangF0ZC21, DBLP:conf/www/WangLF0LC22}, we adopted the widely-used MRR@10, R@10, and R@50 for the passage retrieval task, as evaluation metrics.
Notably, during the testing, we employed an off-the-shelf tool for large-scale vector-similarity search, namely faiss~\cite{johnson2019billion}, to efficiently conduct the relevant document retrieval.  

	

\begin{table}[!t]
\centering
    \caption{Passage retrieval performance~(\%) on the \mbox{TaoLive} dataset in terms of MRR@10, R@10, and R@50. We report the results of two variant methods in different modalities for comparison. 
    }
\begin{tabular}{|l|c|ccc|}
    \hline
    \multicolumn{1}{|c|}{\multirow{2}{*}{\textbf{Model}}} & \multicolumn{1}{c|}{\multirow{2}{*}{\textbf{Modality}}} & \multicolumn{3}{c|}{\textbf{TaoLive}}\\ \cline{3-5} 
     & & MRR@10 & R@10 & R@50 \\ \hline
    OFAR-w/-FixEncoder & All & 19.6 & 33.5 & 43.4 \\ 
    OFAR-w/-FixEncoder & Vision & 3.3 & 6.13 & 11.7 \\ 
    OFAR-w/-FixEncoder & Text & 16.5 & 29.9 & 36.3 \\ \hline
    OFAR-w/o-MaxSim & All & \underline{23.4} & \underline{41.4} & \underline{67.5} \\
    OFAR-w/o-MaxSim & Vision & 4.6 & 7.01 & 18.9 \\
    OFAR-w/o-MaxSim & Text & 21.3 & 38.0 & 62.4 \\ \hline
    OFAR & All & \textbf{37.4} & \textbf{53.2} & \textbf{73.9} \\
    OFAR & Vision & 9.1 & 13.7 & 24.9 \\
    OFAR & Text & 25.5 & 43.2 & 65.9 \\ \hline
    Improvement. $\uparrow$ & - & 59.8\% & 28.5\% & 9.5\% \\ \hline 
\end{tabular}
\label{modelcom}
\end{table}

\subsection{Model Performance}
We design the following two variant methods to justify our model: 1) \textbf{OFAR-w/-FixEncoder}. To show the importance of finetuning the OFA encoders with MaxSim-based contrastive late intersection, this variant uses these OFA encoders to obtain the $\mathbf{E}_q$ and $\mathbf{E}_d$ without training and finetuning and applies faiss in searching related passages during testing. Notably, this variant does not use MaxSim-based contrastive late intersection.     
And 2) \textbf{OFAR-w/o-MaxSim}. To show the benefit of using the MaxSim, we replaced it with the $[CLS]$ token-based similarity score as traditional contrastive loss does. 

Table~\ref{modelcom} shows the performance comparison among different variant methods over \mbox{TaoLive} in terms of MRR@10, R@10, and R@50. The best results are in boldface, while the second best are underlined.  
For a comprehensive comparison, we tested different models with three data configurations: multimodal input (All), pure visual input~(Vision), and pure textual input~(Text). Specifically, All indicates that we used all samples, and Vision (Text) implies that we only used the visual (text) content of each sample. 
From this table, we have the following observation. 
1) Our model OFAR consistently outperforms the other methods in terms of all metrics with significant improvement. This may be due to that our MaxSim-based contrastive late intersection can bring more accurate similarity computation, and largely improve the model performance. 
2) Our OFAR surpasses the OFAR-w/o-MaxSim. This implies the advantage of  MaxSim-based late intersection in query-document similarity computation rather than the simple $[CLS]$ token-based cosine similarity score. 
And 3) only textual modality or visual modality causes a decrease in the model performance. This verifies the importance of taking into account all modalities, including textual modality and visual modality.  

%% file: sec-con.tex
\section{Conclusion}
\label{sec:con}
In this paper, we present a novel multimodal evidence retrieval system, OFAR, based on dual OFA-based encoders, which seamlessly unifies the heterogeneous multimodal data of queries and documents into the same embedding space according to an advanced multimodal pretrained model OFA. 
In order to effectively utilize token-level visual and textual content generated by the OFA encoders, we explore the MaxSim-based late interaction for them. 
To be specific, we improve the conventional triple loss with contrastive loss, where we incorporate more negative documents, to enhance the model’s ability of query-document matching. Finally, we pretrain our model in two widely-used datasets, WebQA and \mbox{MS MARCO}.
And then we finetune it on our internal dataset TaoLive to demonstrate the effectiveness of our method. 